\documentclass[aps,twocolumn,showpacs, showkeys,nofootinbib,floatfix]{revtex4-1}
\usepackage{amsmath}
\usepackage{amsfonts}
\usepackage{txfonts}
\usepackage{graphicx}
\usepackage{dcolumn}
\usepackage{bbm}
\usepackage{amssymb}
\usepackage{latexsym}
\usepackage{titlesec}
\usepackage[colorlinks=true, linkcolor=red, citecolor=blue]{hyperref}
\usepackage{longtable}

\begin{document}

\title{Pursuing the Amplitude of Tensor Mode Power Spectrum in Light of BICEP2}

\author{Baorong Chang}
\email{changbaorong@dlut.edu.cn}

\author{Lixin Xu}
\email{Corresponding author: lxxu@dlut.edu.cn}

\affiliation{Institute of Theoretical Physics, School of Physics \&
Optoelectronic Technology, Dalian University of Technology, Dalian,
116024, P. R. China}

\affiliation{College of Advanced Science \& Technology, 
Dalian University of Technology, Dalian, 116024, P. R. China}

\affiliation{State Key Laboratory of Theoretical Physics, Institute of Theoretical Physics, Chinese Academy of Sciences}

\begin{abstract}
In this brief report, we try to constrain general parameterized forms of scalar and tensor mode power spectra, $P_{s}(k)\equiv A_s(k/k_0)^{n_s-1+\frac{1}{2}\alpha_s\ln(k/k_0)}$ and $P_{t}(k)\equiv A_t(k/k_0)^{n_t+\frac{1}{2}\alpha_t\ln(k/k_0)}$ by the recently released BICEP2 data set plus {\it Planck} 2013, WMAP9 and BAO. We loosen the inflationary consistence relations, and take $A_s$, $n_s$, $A_t$ and $n_t$ as free model parameters, via the Markov chain Monte Carlo method, the interested model parameter space was investigated, we obtained marginalized $68\%$ limits on the interested parameters are: $n_s=0.96339_{-0.00554}^{+0.00560}$, $n_t=1.70490_{-0.56979}^{+0.56104}$, ${\rm{ln}}(10^{10} A_s)=3.08682_{-0.02614}^{+0.02353}$ and ${\rm{ln}}(10^{10} A_t)=3.98376_{-0.54885}^{+0.86045}$. The ratio of the amplitude at the scale $k=0.002 \text{Mpc} ^{-1}$ is $r=0.01655_{-0.01655}^{+0.00011}$ which is consistent with the {\it Planck} 2013 result.
\end{abstract}

%\pacs{95.36.+x, 98.80.Es, 95.35.+d}

%\keywords{accelerated expansion; unified dark fluid} 

\maketitle

\section{Introduction}

The BICEP2 experiment \cite{ref:BICEP21,ref:BICEP22} has detected the B-modes of polarization in the cosmic microwave background. And this observed B-modes power spectrum gives the constraint to the tensor-to-scalar ratio with $r=0.20^{+0.07}_{-0.05}$ at the $1\sigma$ level of the lensed-$\Lambda$CDM model \cite{ref:BICEP21,ref:BICEP22}. And the tensor spectral tilt $n_t$ can be obtained, when the first oder consistency relation, $n_t=-r/8$, was respected. Also, relaxing this consistency relation by taking $n_t$ as a free model parameter \cite{ref:Huang2014}, $r_{0.002}=0.21^{+0.04}_{-0.10}$ and $n_t=-0.06^{+0.25}_{-0.23}$ were obtained by using BICEP2 only. By combining {\it Planck}, WMAP9 and BAO data, it was already found that a blue tilt is slightly favored, but it is still well consistent with flat or red tilt \cite{ref:Wu2014}. However, one can go further by taking generalized parameterized forms of scalar and tensor mode power spectra as  
\begin{eqnarray}
P_{s}(k)&\equiv& A_s(k/k_0)^{n_s-1+\frac{1}{2}\alpha_s\ln(k/k_0)},\label{eq:ps}\\
P_{t}(k)&\equiv& A_t(k/k_0)^{n_t+\frac{1}{2}\alpha_t\ln(k/k_0)},\label{eq:pt}
\end{eqnarray} 
without assuming any idea about inflation, in other words throwing away the consistence relations, just considering the possible deviation from the scale invariant power spectra, i.e. the Harrison-Zel'dovich-Peebles spectra. And how to interpreter it is another issue. Of course, one can relate it to the so-called inflation, where the consistence relations should be respected. And in this way, one can test the viability of inflation models. But, one can also explain it through the bounce expansion. Here $n_s-1$ and $n_t$ are tilts of power spectrum of scalar and tensor modes, $k_0=0.05 \text{Mpc} ^{-1}$ is the pivot scale, $\alpha_s=d n_s/d\ln k$ and $\alpha_t=d n_t/d\ln k$ are the running of the scalar and tensor spectral tilts. The primordial tensor-to-scalar ratio is defined by $r\equiv A_t/A_s$ at different pivot scale, here, they are $r_{0.05}$ defined at $k_0=0.05 \text{Mpc} ^{-1}$ and $r_{0.002}$ defined at $k_0=0.002 \text{Mpc} ^{-1}$. In this paper, without any other specification, $r_{0.002}$ will be donated by $r$. And we also denote $A_t/A_s$ as the amplitude ratio of the tensor and scalar mode power spectrum at $k\equiv k_0$, i.e. the scale independent tensor-to-scalar ratio. In our calculation, adiabatic initial conditions were assumed in this paper. Actually, if one wants to relate the parameterized primordial power spectra, the following relations, the so-called consistency relation should be respected \cite{ref:Mukhanov1999}
\begin{equation}
r=-8c_s n_t,\label{eq:consnt}
\end{equation} 
By taking these parameters, $n_t$ and $A_t$ as free ones, one can test these consistency relation by the recently released BICEP2 data. 

Here, we are mainly focusing on the model parameters which are related to the primordial power spectra. Therefore, in this brief paper, by combing the following data sets, we report the constrained results on the interested parameters:

(i) The newly released BICEP2 CMB B-mode data \cite{ref:BICEP21,ref:BICEP22}.

(ii) The full information of CMB which include the recently released {\it Planck} data sets which include the high-l TT likelihood ({\it CAMSpec}) up to a maximum multipole number of $l_{max}=2500$ from $l=50$, the low-l TT likelihood ({\it lowl}) up to $l=49$ and the low-l TE, EE, BB likelihood up to $l=32$ from WMAP9, the data sets are available on line \cite{ref:Planckdata}.

(iii) For the BAO data points as 'standard ruler', we use the measured ratio of $D_V/r_s$, where $r_s$ is the co-moving sound horizon scale at the recombination epoch, $D_V$ is the 'volume distance' which is defined as
\begin{equation}
D_V(z)=[(1+z)^2D^2_A(z)cz/H(z)]^{1/3},
\end{equation}
where $D_A$ is the angular diameter distance. The BAO data include $D_V(0.106) = 456\pm 27$ [Mpc] from 6dF Galaxy Redshift Survey \cite{ref:BAO6dF}; $D_V(0.35)/r_s = 8.88\pm 0.17$ from SDSS DR7 data \cite{ref:BAOsdssdr7}; $D_V(0.57)/r_s = 13.62\pm 0.22$ from BOSS DR9 data \cite{ref:sdssdr9}. Here the BAO measurements from WiggleZ are not included, as they come from the same galaxy sample as $P(k)$ measurement.

We will present the method and obtained results in the next section \ref{sec:results}. Section \ref{sec:conclusion} is the conclusion.

\section{Constrained Results} \label{sec:results}

To study the effect of the spectral tilt $n_t$ and the amplitude $A_t$ to the B-mode of CMB power spectrum, we use the newest version of {\bf CAMB} \cite{ref:CAMB} code where the the running $\alpha_s$, $\alpha_t$ and amplitude $A_t$ as free mdomel parameters have been included in the forms of the equations (\ref{eq:ps}) and (\ref{eq:pt}). By fixing the other relevant cosmological model parameters but varying $n_t$ or $A_t$, we show the B-mode of CMB power spectrum in Figure \ref{fig:Bmode} for different values of $n_t$ and $A_t$, where the other relevant model parameters were fixed to their values obtained by {\it Planck} 2013 Collaboration group \cite{ref:Planck2013}.

One can clearly see that the variation of values of $A_t$ corresponds to move the B-mode power spectrum along the vertical direction and almost keep the shape untouched. And larger values have large amplitude of the B-mode power spectrum as expected. On the contrast, the values of $n_t$ manages the shapes at the low multipoles $l<150$. Large values of $n_t$ will decrease the amplitude of the B-mode CMB power spectrum. And we should also notice the fact that the effects on the B-mode CMB power spectrum are truely independent on the consistency relations.     

\begin{widetext}
\begin{center}
\begin{figure}[tbh]
\includegraphics[width=8.5cm]{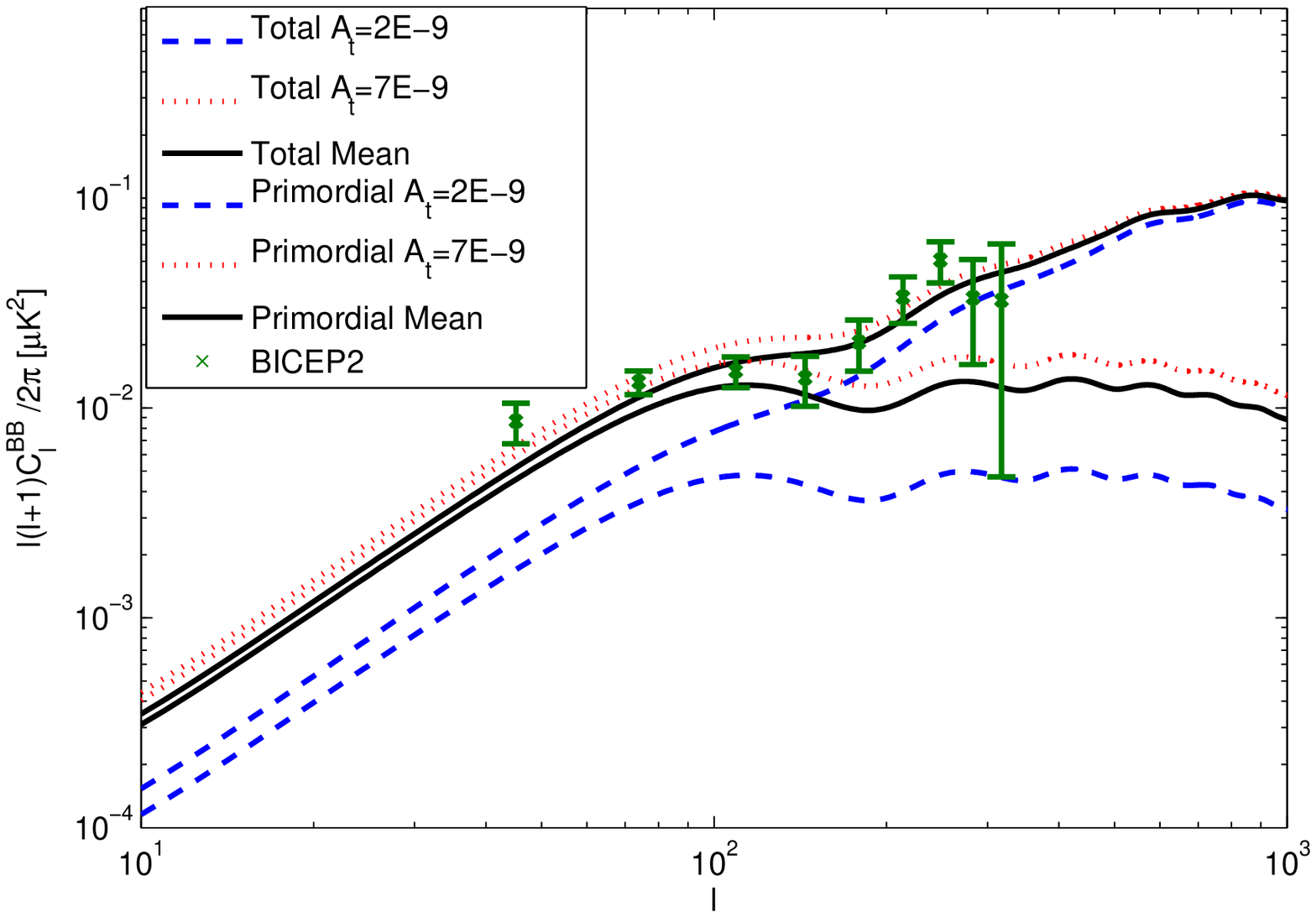}
\includegraphics[width=8.5cm]{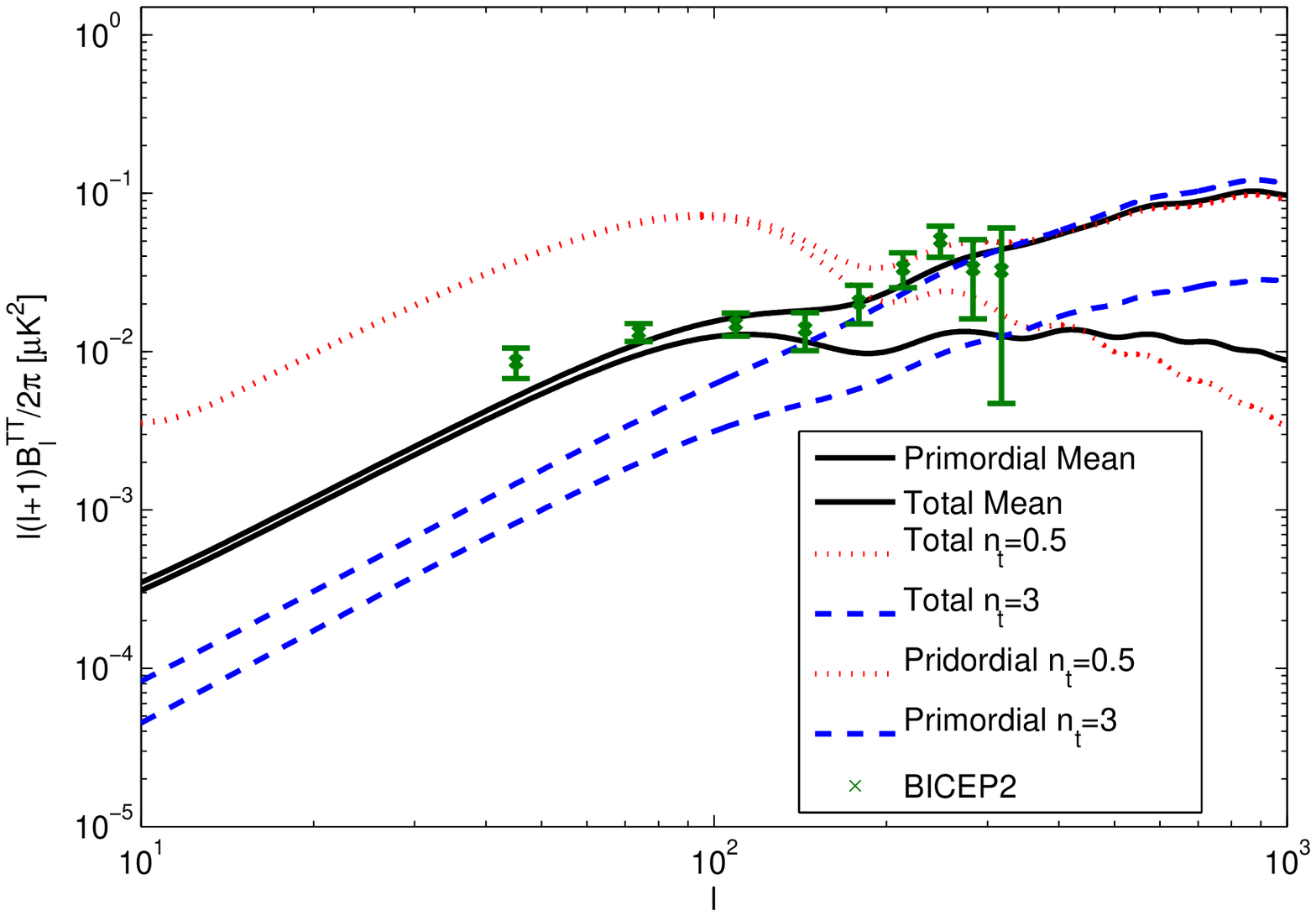}
\caption{The simulated BB power spectrum for varying $n_t$ or $A_t$ where the other relevant model parameters are fixed to their best fit values obtained by {\it Planck} 2013 Collaboration group \cite{ref:Planck2013}. The left panel shows the B-mode CMB power spectrum for different values of $A_t=2\times 10^{-9}, 3.98376\times 10^{-9}, 7\times 10^{-9}$ respectively for the total lensed and primordial cases. The right panel is same as the left panel, but for varying $n_t=0.5,1.70490, 3$ respectively.}\label{fig:Bmode}
\end{figure}
\end{center}
\end{widetext}

To investigate the model parameter space,
\begin{equation}
P=\{\Omega_b h^2,\Omega_c h^2,  100\theta_{MC}, \tau, n_s, n_t, {\rm{ln}}(10^{10} A_s),{\rm{ln}}(10^{10} A_t)\},
\end{equation}
we performing a global fitting on the {\it Computing Cluster for Cosmos} by using the publicly available package {\bf CosmoMC} \cite{ref:MCMC}, which includes {\bf CAMB} \cite{ref:CAMB} to calculate the CMB power spectra that has been used to produce the BB power spectrum in Figure \ref{fig:Bmode}. The running was stopped when the Gelman \& Rubin $R-1$ parameter $R-1 \sim 0.01$ was arrived; that guarantees the accurate confidence limits. We set the {\bf inflation\_consistency = F} in the input {\bf params\_CMB\_defaults.ini} file. The obtained results are shown in Table \ref{tab:results}. The obtained contour plots are shown in Figure \ref{fig:contour}. 

%\begin{widetext}
\begingroup                                                                                                                     
%\squeezetable                                                                                                                   
\begin{center}                                                                                                                  
\begin{table}[tbh]                                                                                                                   
\begin{tabular}{cccc}                                                                                                            
\hline\hline                                                                                                                    
Parameters & Priors & Mean with errors & Best fit \\ \hline
$\Omega_b h^2$ & $[0.005,0.1]$ & $0.02212_{-0.00024}^{+0.00024}$ & $0.02233$\\
$\Omega_c h^2$ & $[0.001,0.99]$ & $0.1185_{-0.0017}^{+0.0017}$ & $0.1188$\\
$100\theta_{MC}$ & $[0.5,10]$ & $1.04143_{-0.00056}^{+0.00057}$ & $1.04158$\\
$\tau$ & $[0.01,0.81]$ & $0.09060_{-0.01352}^{+0.01184}$ & $0.09739$\\
$n_s$ & $[0.9,1.1]$ & $0.96339_{-0.00554}^{+0.00560}$ & $0.96497$\\
$n_t$ & $[-5,5]$ & $1.70490_{-0.56979}^{+0.56104}$ & $1.37722$\\
${\rm{ln}}(10^{10} A_s)$ & $[2.7,4]$ & $3.08682_{-0.02614}^{+0.02353}$ & $3.10357$\\
${\rm{ln}}(10^{10} A_t)$ & $[0.01,10]$ & $3.98376_{-0.54885}^{+0.86045}$ & $3.66014$\\
\hline
$\Omega_\Lambda$ & $...$ & $0.69292_{-0.01019}^{+0.01016}$ & $0.69286$\\
$\Omega_m$ & $...$ & $0.30708_{-0.01016}^{+0.01019}$ & $0.30714$\\
$z_{re}$ & $...$ & $11.08681_{-1.05317}^{+1.06072}$ & $11.63177$\\
$H_0$ & $...$ & $67.84886_{-0.77032}^{+0.76690}$ & $67.95166$\\
$r$ & $...$ & $0.01655_{-0.01655}^{+0.00011}$ & $0.01851$\\
$A_t/A_s$ & $...$ & $3.06406_{-2.55488}^{+0.97924}$ & $1.74468$\\
${\rm{Age}}/{\rm{Gyr}}$ & $...$ & $13.80101_{-0.03741}^{+0.03730}$ & $13.77666$\\
\hline\hline                                                                                                                    
\end{tabular}                                                                                                                                                                                                                                  
\caption{The mean values with $1\sigma$ errors and the best fit values of the model parameters and derived cosmological parameters, where the {\it Planck} 2013, WMAP9 and BICEP2 data sets were used.}\label{tab:results}                                                                                                 
\end{table}                                                                                                                     
\end{center}                                                                                                                    
\endgroup   
%\end{widetext}

\begin{widetext}
\begin{center}
\begin{figure}[tbh]
\includegraphics[width=15cm]{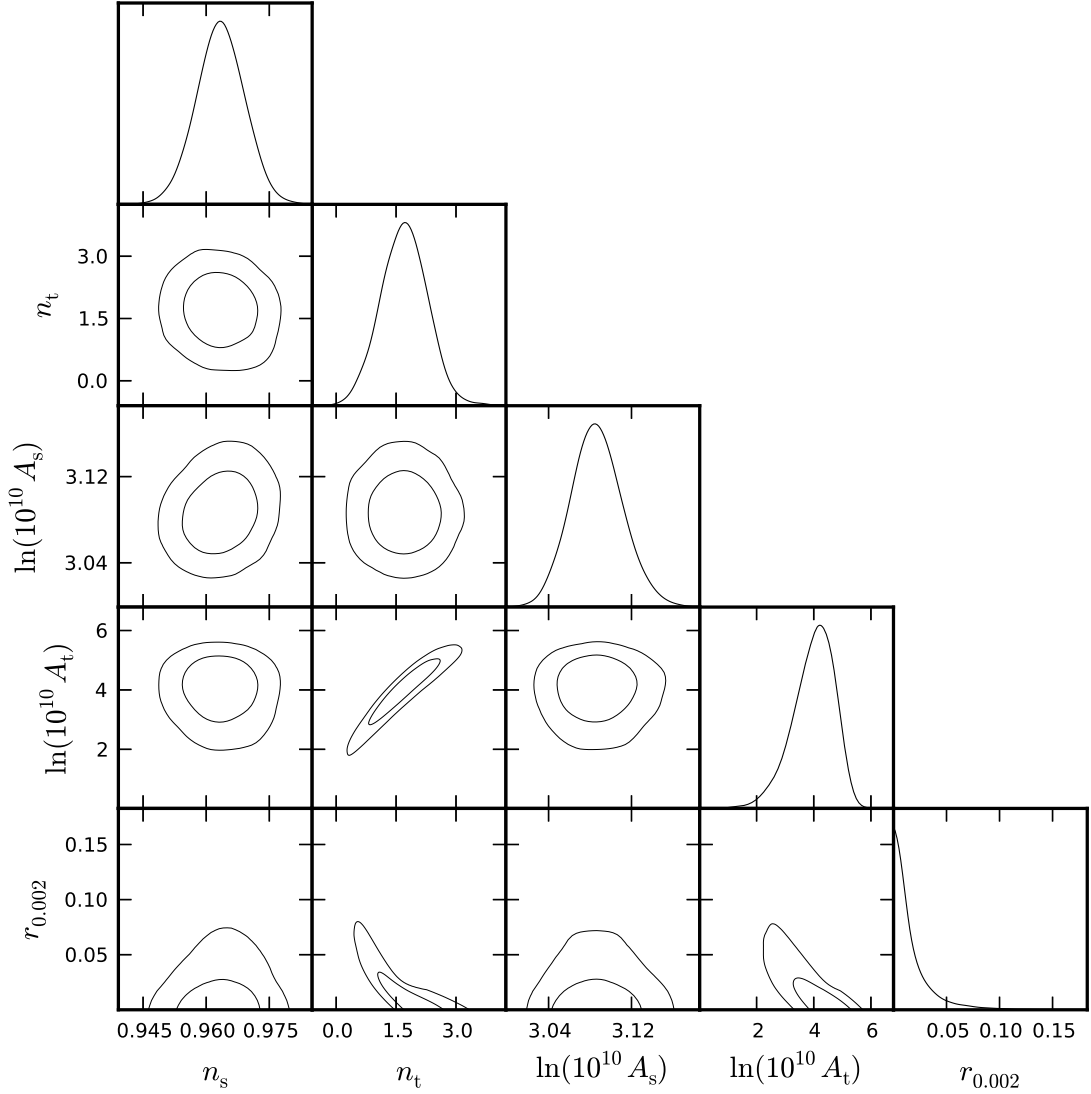}
\caption{The 1D marginalized distribution and 2D contours for interested model parameters with $68\%$ C.L., $95\%$ C.L. by using  {\it Planck} 2013, WMAP9 and BICEP2.}\label{fig:contour}
\end{figure}
\end{center}
\end{widetext}

\section{Conclusion} \label{sec:conclusion} 

In this brief paper, we loosen the inflation consistency relation constraint, and take the spectral tilts $n_s$, $n_t$, $A_s$ and $A_t$ as free model parameters. Combining the recently released BICEP2 data, {\it Planck} 2013, WMAP9 and BAO via the MCMC method, the model parameter space was scanned. We found that $n_s=0.96339_{-0.00554}^{+0.00560}$, $n_t=1.70490_{-0.56979}^{+0.56104}$, ${\rm{ln}}(10^{10} A_s)=3.08682_{-0.02614}^{+0.02353}$ and ${\rm{ln}}(10^{10} A_t)=3.98376_{-0.54885}^{+0.86045}$. The ratio of the amplitude at the scale $k=0.002 \text{Mpc} ^{-1}$ is $r=0.01655_{-0.01655}^{+0.00011}$ which is consistent with the {\it Planck} 2013 result. And a blue tensor tilt is favored at $1\sigma$ C.L.. And $n_t$ is positive above $2\sigma$ C.L.. It implies the broken of consistency relation $r=-8c_s n_t$ at $2\sigma$ C.L., when the speed of sound $c_s>0$ is respected.

\acknowledgements{This work is supported in part by NSFC under the Grants No. 11275035 and "the Fundamental Research Funds for the Central Universities" under the Grants No. DUT13LK01.}

\end{document}